\documentclass[conference]{IEEEtran}
\IEEEoverridecommandlockouts
\usepackage{cite}
\usepackage{amsmath,amssymb,amsfonts}
\usepackage{algorithmic}
\usepackage{graphicx}
\usepackage{textcomp}
\usepackage{xcolor}
\usepackage{hyperref}
\usepackage{tablefootnote}
\usepackage{threeparttable}
\usepackage{booktabs}
\def\BibTeX{{\rm B\kern-.05em{\sc i\kern-.025em b}\kern-.08em
    T\kern-.1667em\lower.7ex\hbox{E}\kern-.125emX}}
\begin{document}
\IEEEoverridecommandlockouts
\IEEEpubid{\makebox[\columnwidth]{ 979-8-3503-7608-1/24\$31.00 \copyright2024 IEEE \hfill} \hspace{\columnsep}\makebox[\columnwidth]{ }}
\title{Optimizing High-Level Synthesis Designs with
Retrieval-Augmented Large Language Models\\
}

\author{\IEEEauthorblockN{Haocheng Xu}
\IEEEauthorblockA{\textit{Dept. of EECS} \\
\textit{University of California, Irvine}\\
Irvine, USA \\
haochx5@uci.edu}
\and
\IEEEauthorblockN{Haotian Hu}
\IEEEauthorblockA{\textit{School of the Gifted Young} \\
\textit{University of Science and Technology of China}\\
Hefei, China \\
haotianhu@mail.ustc.edu.cn}
\and
\IEEEauthorblockN{Sitao Huang}
\IEEEauthorblockA{\textit{Dept. of EECS} \\
\textit{University of California, Irvine}\\
Irvine, USA \\
sitaoh@uci.edu}
}

\maketitle

\begin{abstract}

High-level synthesis (HLS) allows hardware designers to create hardware designs with high-level programming languages like C/C++/OpenCL, which greatly improves hardware design productivity. However, existing HLS flows require programmers' hardware design expertise and rely on programmers' manual code transformations and directive annotations to guide compiler optimizations. Optimizing HLS designs requires non-trivial HLS expertise and tedious iterative process in HLS code optimization. Automating HLS code optimizations has become a burning need. 
Recently, large language models (LLMs) trained on massive code and programming tasks have demonstrated remarkable proficiency in comprehending code, showing the ability to handle domain-specific programming queries directly without labor-intensive fine-tuning. 
In this work, we propose a novel retrieval-augmented LLM-based approach to effectively optimize high-level synthesis (HLS) programs.  
Our proposed method leverages few-shot learning, enabling large language models to adopt domain-specific knowledge through natural language prompts. 
We propose a unique framework, Retrieve Augmented Large Language Model Aided Design (RALAD), designed to enhance LLMs' performance in HLS code optimization tasks. RALAD employs advanced embedding techniques and top-\emph{k} search algorithms to dynamically source relevant knowledge from extensive databases, thereby providing contextually appropriate responses to complex programming queries. Our implementation of RALAD on two specialized domains, utilizing comparatively smaller language models, achieves an impressive 80\% success rate in compilation tasks and outperforms general LLMs by 3.7 -- 19$\times$ in latency improvement. 
\end{abstract}


\section{Introduction}

The last decades have witnessed the emergence of domain-specific accelerators in various application domains, which greatly improves the computational efficiency of computer systems. In the field of domain-specific accelerator design, high-level synthesis (HLS) has become a popular hardware design method as it allows designers to write C/C++/OpenCL code to generate hardware designs with the help of HLS compilers. Even though HLS greatly improves hardware design efficiency compared to RTL-level design flows, existing HLS compilers heavily rely on programmers' manual code transformation and compiler directive insertion to guide HLS compiler optimizations, making the HLS optimization time-consuming and sub-optimal. Optimizing HLS effectively and automatically has been a big challenge. In this work, we propose a novel generative AI-based solution to this HLS optimization challenge, \textbf{RALAD}, a hybrid approach that combines large language models (LLMs) based code generation with HLS domain-specific optimization knowledge base, enabled by retrieval-augmented generation (RAG) technology.  
\begin{figure}
    \centering
    \includegraphics[width=0.5\textwidth]{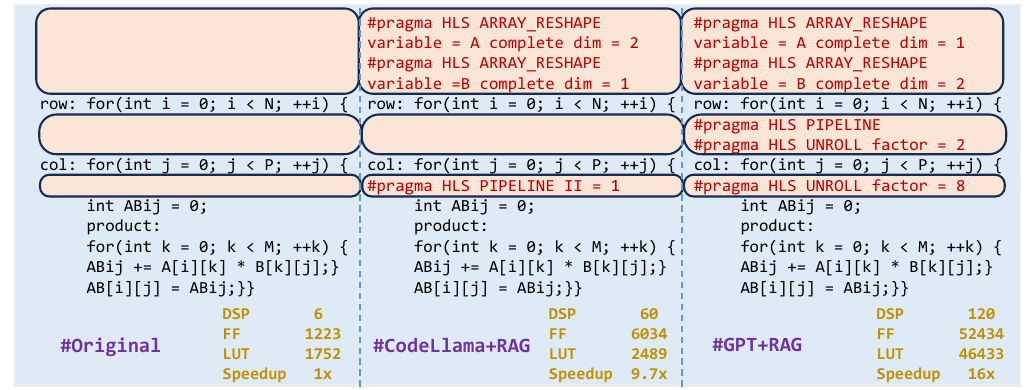}
    \caption{(left) Original HLS C code; (middle)  Code-Llama-13B + RAG generated HLS C code; (right)  GPT-3.5 + RAG generated HLS C code. }
    \label{fig:compare}
\end{figure}

Large language models (LLMs) have demonstrated exceptional performance across numerous domains and hold immense potential in fields like coding, medicine, and law, closely approaching human-level expertise \cite{naveed2024comprehensive}. However, they still suffer when answering questions that require knowledge of local conditions, specific varieties, or up-to-date data\cite{b1}. One way to address this issue is to fine-tune the pre-trained LLMs on specific datasets to adapt the domain knowledge. However, this approach usually requires retraining on partial or full layers of the pre-trained LLMs on a specific dataset, which is compute-intensive and time-consuming. Moreover, fine-tuning will be required each time when the LLMs need to be deployed in a different domain. Another way to address the domain knowledge issue is through \emph{retrieval-augmented generation (RAG)} \cite{b3}. RAG allows LLMs to retrieve information from an external source, augment their input with relevant context, and generate more informed, accurate prompts to guide the LLMs' responses to domain-specific queries. 

Code optimization, for instance, can potentially benefit from RAG by fetching the relevant information from previous similar code optimization examples and using them to guide new code optimizations or perform debugging. At high level, our proposed flow works as follows. The flow takes users' input of a piece of code with the optimization instructions. Those original prompts will be embedded by one of the pre-trained embedding models; then, a query index search will try to match the prompts embedding with the relevant document (code) embedding in the customized code base; carefully crafted prompts will be generated by integrating the instructions, original prompts, and retrieved code; finally, the pre-trained LLMs will process the crafted prompts to optimize the code. This RAG mechanism for code-generation systems does not require training or fine-tuning of the existing LLMs. 

\section{Methodology}
\begin{figure*}
    \centering
    \includegraphics[width = \textwidth]{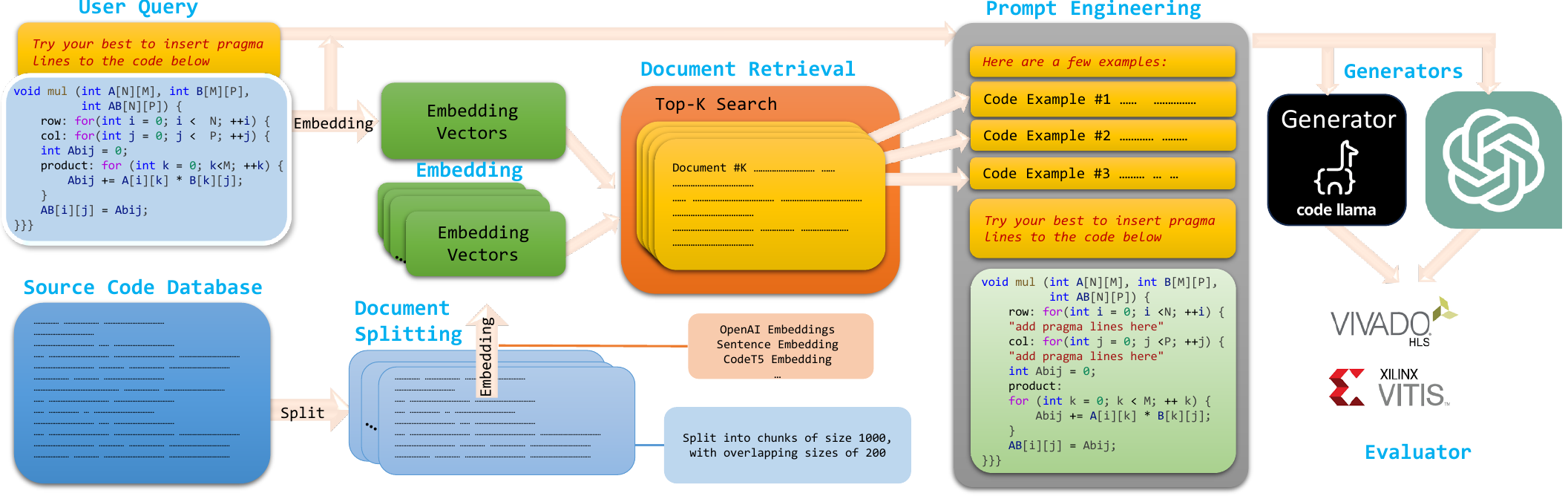}
    \caption{The Overview of Our Proposed RALAD HLS Code Optimization Framework}
    \label{fig:framework}
\end{figure*}

Fig. \ref{fig:framework} shows the basic structure of our RALAD code optimization framework which includes source code splitting, embedding, related code search, prompt reconstruction, and code generation. We will introduce each part in the following subsections.

\subsection{Document Embedding}

Prior works have demonstrated the promising performance boost that RAG can bring to text generation. Patrick et al. \cite{b3} explored the application of RAG on knowledge-intensive tasks using Wikipedia as their primary data source. Fu et al. \cite{b5} leverage the most relevant code demonstration as references for processing the query code (QC). Inspired by prior works, we show that retrieving similar text or code from programming books or guides (in natural language), specifically those focusing on the HLS domain could be beneficial for our intended tasks. 

Given the highly limited amount of source code datasets that come with HLS pragma annotations, we use ``Parallel Programming for FPGAs'' (\emph{pp4f}) book \cite{b4} as our foundational dataset due to its potential to offer instructive examples and relevant documentation for learning. To preserve the integrity of pragma directives embedded within the source code, each source file was segmented into chunks to ensure the comprehensive retention of pragma-related information, which is critical for understanding and applying HLS optimizations. 

Following the segmentation of source code into chunks, we explored various embedding techniques including, OpenAI embedding, sentence embedding, and Code T5 embedding. Embeddings are essentially vectors generated by models to encapsulate significant data within contexts to enhance our framework's comprehension of the source code's semantics and structure while reducing the search complexity. 

More specifically, we divided the entire \emph{pp4f} book into 699 chunks, each with a chunk size of 1000 characters and with an overlap size of 200 characters between consecutive segments. This overlapping strategy was employed to minimize the risk of dividing the wrong part of the codes into different chunks to ensure the integrity of the code examples within the individual chunks. 

Similarly, the queries from the users were also transformed into vector representation, to ensure a direct comparison and retrieval process based on the semantic similarity between the query code and the source code.


\subsection{Document Retrieval}
Identifying the top-K results for our query can be efficiently achieved using specialized libraries, such as FAISS \cite{b6}, or through a direct comparison of distances between the encoded query and the encoded code chunks, selecting the top K most similar. Zhou et al.\cite{zhou23docprompting} declared that retrieving the top 4 relevant documents is generally sufficient for most applications as increasing the retrieved documents leads to larger memory consumption without proportional benefits. 

\subsection{Prompt engineering}
Noor et al. \cite{b9} investigated how to build few-shot learning prompts for code-processing tasks. Inspired by them, we have developed a version of prompt design based on their structure. 

\begin{figure}
    \centering
    \includegraphics[width = 0.38\textwidth]{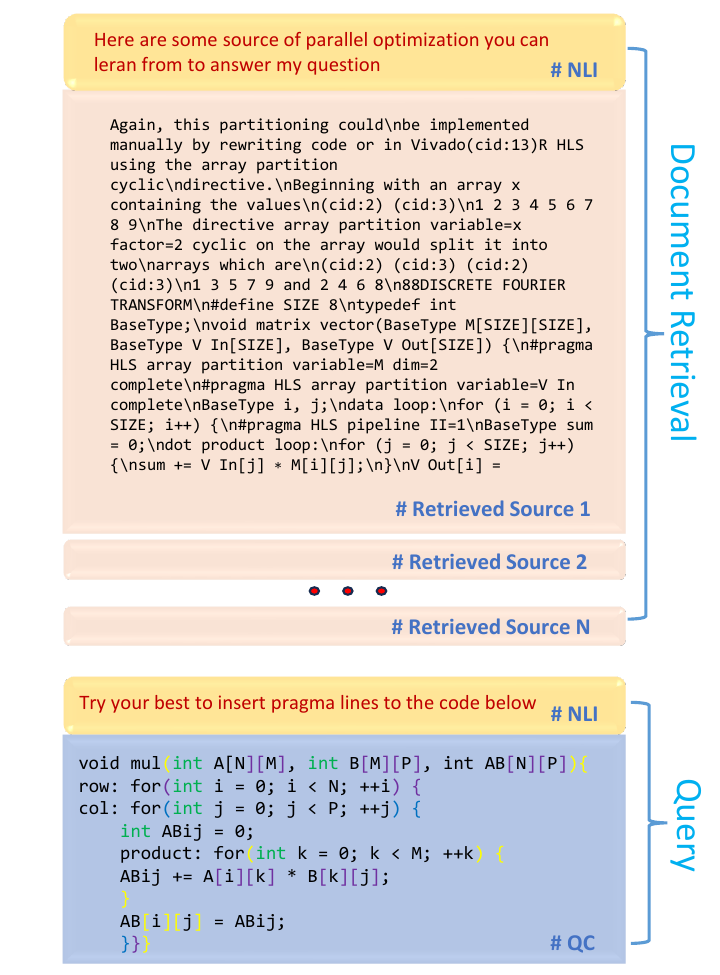}
    \caption{The Structure of Input Prompts}
    \label{fig:2}
\end{figure}

For HLS Optimization tasks, we construct prompts by integrating three key components: (a) Natural language instructions (NLI) (b) Retrieved Documents (RD) (c) Query with to-be-optimized code (QC). To demonstrate, we use the example of optimizing Matrix Multiplication in High-Level Synthesis, combining the above elements to form a final prompt, as shown in Fig. \ref{fig:2}.

\begin{figure}
    \centering
    \includegraphics[width = 0.45\textwidth]{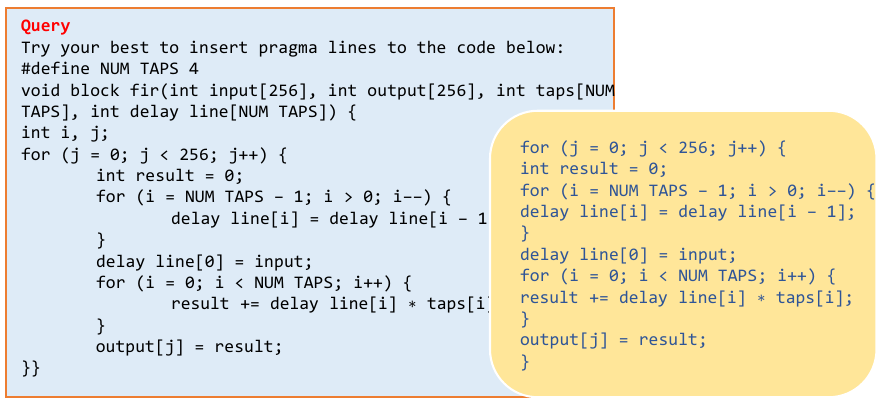}
    \caption{We ignore the initialization part of the query code, only extract the main loops as input, as shown in the yellow part of the graph}
    \label{fig:3}
\end{figure}

To refine the prompts, we introduce additional features. Taking optimizing an FIR Filter as an example, illustrated in Fig. \ref{fig:3}, we focus on the code's main body, ignoring the initialization part of the function. This decision is driven by the observation that during document retrieval, variable names can obscure critical information, leading to model outputs that are irrelevant to the query.

\begin{figure}
    \centering
    \includegraphics[width = 0.45\textwidth]{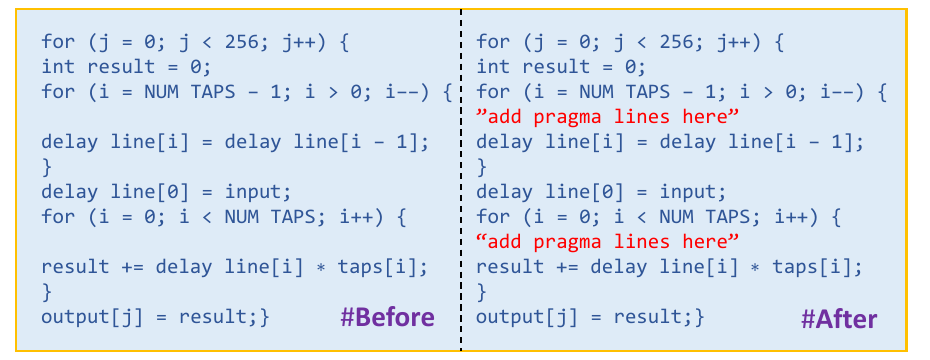}
    \caption{Manual Annotations: We add hints to the prompt manually by human experts to enhance the performance of the generator}
    \label{fig:4}
\end{figure}

In addition, we apply the expert intervention to manually insert ``\texttt{add pragma lines here}'' annotations in our query at positions where pragma lines could be expected, as shown in Fig. \ref{fig:4}. This step aims to improve the model's capability to generate HLS synthesizable codes by providing explicit hints about optimization locations. The effectiveness of this technique is further evaluated in the experiment sections.
\begin{table}[htbp]
\caption{Percentage of Synthesizable Cases for Models with Various Settings (the higher the better)}
\begin{center}
\begin{tabular}{l|c|c|c}
\hline
\textbf{Valid Percentage} & \textbf{Zero Shot}& \textbf{RAG}& \textbf{RAG + Annotations} \\
\hline
Code Llama-7B model & 0\% & 30\% & 50\%\\
\hline
Code Llama-13B model & 10\% & 60\%&80\%\\
\hline
\end{tabular}
\label{tab1}
\end{center}
\end{table}
\subsection{Generator}

For the generator component of our framework, we explored various LLMs including GPT-3.5 and other fine-tuned open-sourced LLMs for code tasks, like CodeLlama \cite{b10} and T5 Code \cite{b11}. We selected CodeLlama \cite{b10}, a variant of Llama 2 specifically fine-tuned for code-related tasks. CodeLlama was developed by extending the training of Llama 2 with additional code-specific datasets, thereby enriching its understanding of programming languages and code optimization tasks. This choice was motivated by the necessity for our model to acquire a foundational comprehension of the code we aim to optimize. 

Although the T5 code model \cite{b11}, known for its encoder-decoder architecture, was also considered, it failed to produce valuable outputs for our specific use case. Even when supplemented with RAG, T5 model's performance did not improve significantly. This outcome led us to conclude that effective code optimization requires more than just pattern recognition; T5 model's lack of intrinsic understanding of C code proved to be a significant barrier, preventing it from generating useful contributions to HLS code optimization tasks.





\subsection{Benchmarks}
In our study, we benchmarked the capabilities of GPT-3.5 and the open-sourced LLMs, CodeLlama-7b and CodeLlama-13b, specifically in the context of HLS pragma insertion. Aligning with the methodologies outlined in previous research, we focused on a fundamental task in HLS optimization: Matrix Multiplication. To assess the performance of these models, we employed the Pass @k metric (measuring the successful compilations in k attempts on the same example) with or without RAG. This approach allowed us to systematically compare the effectiveness of each model before and after the application of RAG, as well as against the performances of alternative models, providing a comprehensive overview of their respective capabilities in optimizing code for HLS tasks.

\begin{table}[htbp]
\caption{Latency Evaluation of Output Design With RAG}
\begin{center}
\begin{tabular}{l|r|r|r|r|r}
\toprule
\textbf{MatMul} & \textbf{DSP}& \textbf{FF} & \textbf{LUT} & \textbf{Latency} & \textbf{Speed up}\\
\midrule 
Original HLS code & 6 & 1223 & 1752 & 30.08 $\mu$s & 1.00$\times$\\
\hline
GPT-3.5 Raw & 3 & 616 & 512 &60.07 $\mu$s&0.51$\times$\\
\hline
C-Llama-13B Raw* & -- & -- & -- & -- & -- \\
\hline
C-Llama-13B + RAG & 60  & 6034 & 2489 & 3.09 $\mu$s& 9.74$\times$\\
\hline
GPT-3.5 + RAG & 120& 52434 & 46433 & \textbf{1.88} $\mu$s & \textbf{16.00$\times$}\\
\toprule
\textbf{SpMM} & \textbf{DSP} & \textbf{FF} & \textbf{LUT} & \textbf{Latency} & \textbf{Speed up}\\
\midrule
Original code & 3 & 583 & 390 & 2.41 $\mu$s & 1.00$\times$\\
\hline
GPT-3.5 Raw&3 & 506 & 351 & 8.71 $\mu$s & 0.28$\times$\\
\hline
C-Llama-13B + RAG & 6 & 1550 & 1722 & 2.31 $\mu$s &1.04$\times$\\
\hline
GPT-3.5 + RAG & 6 & 1401 & 1133 & \textbf{2.11} $\mu$s & \textbf{1.14}$\times$ \\
\hline
\end{tabular}
\label{tab2}
\begin{tablenotes}\footnotesize
 * Code Llama-13B Raw did not generate synthesizable code; \\  \hspace{1mm} ``C-Llama'' stands for Code Llama
\end{tablenotes}
\end{center}
\end{table}

\section{Experiment}

In this section, we detail our experimental approach and evaluate the efficacy of RAG in optimizing pragma directives. Furthermore, we explore specific instances where our methodology surpasses the performance of zero-shot models.
\subsection{Experiment Design}

Our process began with encoding ``Parallel Programming for FPGAs'' by Ryan et al. \cite{b4} into vector representations. Subsequently, we utilized the FAISS library \cite{b6} to construct an efficient index for these vectors. Guided by insights from Ma et al.\cite{b8}, who demonstrated that the top 4 retrieved documents yield optimal results for Large Language Models, we adapt to this recommendation by selecting the four most similar pieces of code to craft our prompts.

In our experiment, we measure the Average Case Real-Time Latency of the outputs generated by our method using RAG and zero-shot approaches, employing the Vivado HLS tool for evaluation.
We specifically target a series of examples in HLS to gauge the performance of our technique. Various methodologies have been proposed for evaluating model output performance. For instance, Fu et al. \cite{b5} utilize the Pass @k metric, which quantifies the number of successful compilations within k attempts for the same example in addition to the latency. This dual approach allows us to comprehensively understand the effectiveness of our method in optimizing code for HLS tasks. We also report the resource utilization from the Vivado HLS tool.

\subsection{Result Analysis}

As shown in TABLE \ref{tab1}, we present a comparison of outcomes before and after RAG across 10 attempts. The data reveal a significant increase in the successful compilation pass rate upon integrating retrieved documents. Our method improves the pass@10 rate of both the CodeLlama-7b and CodeLlama-13b models by 40\% and 50\%, respectively, in the FIR filter baseline example. Additionally, the incorporation of supplementary hints by human experts into the query further increases the pass@10 rate of our model's output, proving the benefits of our refined methodology.

We benchmarked our outputs using standard examples specific to the HLS domain to evaluate their effectiveness. As shown in TABLE \ref{tab2}, our results indicate that the CodeLlama-13b model with RAG achieved 9.737$\times$ speed up in the Matrix Multiplication task. In comparison, GPT-3.5 with RAG reached 16$\times$ speedup. However, the GPT-3.5 demands significantly more resources, including double the Digital Signal Processing (DSP) usage, 8$\times$ increase in Flip-Flops (FF), and an 18$\times$ rise in Look-Up Tables (LUT). This substantial resource consumption raises critical questions regarding the validity of GPT-3.5's optimization when trading resources for a comparatively modest performance improvement. In the Sparse Matrix Multiplication task, both Code Llama-13b and GPT-3.5 with RAG performed closely, utilizing a similar amount of resources.


\section{Conclusion}
We proposed a unique framework, Retrieve Augmented Large Language Model Aided Design (RALAD), engineered to enhance LLMs' performance in HLS tasks. Our implementation of RALAD on two specialized domains, utilizing comparatively smaller language models, has yielded promising results. 
\section{Related Work and Future Work}
Our framework, RALAD, represents an initial exploration of the design space of code optimization in High-Level Synthesis with large language models. There are some previous work targeting LLM-aided code generation tasks, especially for language with extensive pre-training resources like Python\cite{chen2024improving,nijkamp2023codegen}.


\textbf{Generator Selection} 
In our work, we only tested the limited LLMs for the generator. The performance difference between CodeLlama and T5 Code led us to conjecture that a model must first acquire some foundational understanding of coding principles to effectively address domain-specific challenges in HLS. To overcome this, some fine-tuning with the HLS dataset could be beneficial. Moreover, there are some other code-gen LLMs like GitHub Copilot\cite{codepilot} and Codex\cite{Codex}.

\textbf{Source Retrieving}
Documents and datasets with higher quality in the target domain are needed to retrieve more precise and efficient materials. Sometimes, getting those correct documents could be hard. We may also sometimes retrieve irrelevant or inaccurate documents from the datasets and perform knowledge correction via different types of documents retrieved before generation \cite{yan2024corrective}. 
To deal with this, more processes are required to manipulate our prompt.

\textbf{Other Possibilities}
Chain-of-thought prompting \cite{shypula2023learning} will iteratively increase the performance of the output,  proving especially beneficial when multiple optimization layers are employed. Performance-conditioned generation, on the other hand,  requires different implementations of identical examples to further distinguish the performance difference across the same or similar optimization stacks, thus enabling the most substantial performance improvements.

\bibliographystyle{unsrt}
\bibliography{references}
\vspace{12pt}

\end{document}